# Micromachining structured optical fibres using focused ion beam (FIB) milling

**Cicero Martelli**

Optical Fibre Technology Centre, School of Chemistry, University of Sydney, 206 National Innovation Centre, Eveleigh 1430, Sydney, NSW, Australia, and

School of Electrical and Information Engineering, University of Sydney, NSW 2006, Australia

**Paolo Olivero**

School of Physics, Microanalytical Research Centre, The University of Melbourne, Parkville, Victoria 3010, Australia

**John Canning**

Optical Fibre Technology Centre, School of Chemistry, University of Sydney, and

**Nathaniel Groothoff**

Optical Fibre Technology Centre, School of Chemistry, University of Sydney, and

School of Physics, University of Sydney, NSW 2006, Australia

**Brant Gibson and Shane Huntington**

Quantum Communications Victoria, University of Melbourne, VIC , Australia

A focused ion beam is used to mill side holes in air-silica structured fibres. By way of example, side holes are introduced in two types of air-structured fibres (1) a photonic crystal four-ring fibre and (2) a 6-hole single ring step index structured fibre.

OCIS codes:                                                                                       .

The ability to insert gases, liquids, metallic wires and other functional materials, both prior to and subsequent to insertion of the fibre into a system, is crucial to many applications of air structured fibres such as photonic crystal fibres[1] and Fresnel fibres[2,3]. For example, acetylene gas is used in wavelength reference cells for telecommunication devices[4] and liquid crystals for optical switching and signal modulation[5]. Micro volume cells for optical constants characterization have also been demonstrated[6] and can benefit from the possibility of having a constant flow of material through the sensing media. Consequently, there is interest in accessing the channels from the side. To date two main methods have been employed: (1) direct 193 nm laser ablation[7] and (2) femtosecond laser processing followed by etching[8]. 193nm laser ablation using long pulses leads to spreading of the coupled energy beyond the interaction area, which generally limits the resolution of the hole sizes obtained. Femtosecond laser processing overcomes these issues by ionizing and ablating material on a timescale shorter than the phonon-decoupling rate thus limiting thermal contributions to the mechanism. However, it has the disadvantage of requiring additional post etching to remove material inside the hole.

In this letter, we describe an alternative approach based on ion beam processing. Focused ion beams (FIB) have been used in a number of fibre based applications including modification of waveguide properties[9] fabrication of long period gratings[10], micromachining of the fibre end tip for atomic force microscopy[11] and the precision cutting of photonic crystal fibres[12]. It has also been used to write gratings in silicon optical waveguides[13]. As a consequence of applications in semiconductor lithography, the technique nowadays is mature and dual beam (SEM-FIB) instruments are commercially available. Here, we use a commercial FIB system to create access holes into structured fibres with micron spatial resolution. This method has the advantage of allowing machining with a spatial resolution comparable to that obtained with femtosecond technology but in some cases without the requirements of additional etching steps. Two types of structured fibres are used in the experiments reported here: *fibre (1)*, a photonic crystal fibre with a solid core surrounded by four rings of holes, and *fibre (2)*, a simple step index core fibre surrounded by six large holes. Figure 1 shows the cross-section of these two structured fibres taken with an optical microscope. The central core of *fibre (2)* has germanium making this fibre photosensitive to standard UV writing technologies at 244 nm. These fibres have been used to demonstrate novel measurement techniques – for example, the zeta potential of a microfluidics channel, measuring the streaming current generated as a function of flow[14].

The side holes can be engineered to interconnect the internal micro channels in a single fibre or to allow interconnection of multiple fibres using other thin capillaries. This is of particular importance – such channels enable the integration of the functionalities of various components now available that allow the development of what has been recently proposed by Canning: "Lab-in-a-fibre" technology[15]. These ingredients include engineered channels, gratings, thin films[16], and attenuators and etalons[17].

FIB milling is based on energy transfer of accelerated ions to chemically bonded atoms. The collision transfers sufficient energy to overcome the atom-atom binding energy and the material work function leading to ejection of the material. This process is called sputtering. During sputtering, a portion of the ejected atoms or molecules are re-deposited on the exposed region making it difficult to control the amount of material removed. The milling rate for a given material depends on the ion energy and species, the angle of incidence and the surrounding atmosphere[18]. A Quanta 200 3D (FEI Company) focused ion beam ($Ga^+$ ion source) was used to mill holes from the side in the fibre samples. The acceleration of positive ions into the sample induces the emission of backscattered and sputtered ions as well as secondary electrons with varying rates. These electrons only partially compensate the net injection of positive $Ga^+$ ions. Therefore overall there is excess charge building upon the surface of the material. This can cause a random deflection of the incident beam over time. The problem is effectively solved by covering the sample with conducting material to dissipate the excess electric charge to/from the sample holder. Practically, the samples are fixed with conductive carbon tape to the grounded sample holder, and the micromachining is performed as close as possible (i.e. ~1 mm) to the contact point. Under such conditions, no severe surface charging is observed that could represent an obstacle to fabrication with micron spatial resolution. In order to obtain a high milling rate, a high ion current (20 nA) is used which determines a relatively large beam spot size (~0.5 µm). Higher spatial resolution can be achieved at lower ion currents (down to 1-10 nm) but in such conditions the milling time is significantly increased.

*Fibre (1):* The photonic crystal fibre ($\phi_{external}$ ~ 100 µm, $\phi_{average\ cladding\ hole}$ ~ 4.5 µm, pitch between holes $\Lambda$ ~ 8.5 µm ) is etched using hydrofluoric acid (HF) to reduce the amount of surrounding silica the ion beam later needs to penetrate before accessing the lattice structure. This assists in reducing the milling time by ~10 minutes. Four different fibre samples of *fibre (1)* are fixed to sample holders, without control of their circular orientation, in order to analyze the impact of internal holes on the milling depth. Holes with $\phi = 10$ µm and depths ranging over (10 - 70) µm were milled.

Figure 2 shows the optical microscope side-view images of the fabricated holes. Although the resolution of the measurement is affected by the presence of the air holes and the curvature of the fibre, deep holes were milled. The measured depths against the milling times are plotted in Figure 3. An estimated sputtering yield for the photonic crystal fibre is determined considering a linear approximation of the hole depth as function of the milling time. It should be noted that between zero and five microns of depth this dependence is unlikely to be linear since the surface charges on the silica fibre samples lead to deflection of the ion beam and therefore low sputtering. After some milling time and material removal, a new charge equilibrium is reached and the milling, within experimental error, is approximately linear. The amount of sputtered molecules as function of the number of $Ga^+$ ions bombarding the material is plotted in Figure 3 for both solid silica and the PCF fibre. The former has a sputtering yield of 1.04 +/- 0.21 while for solid silica it is 0.85 [19]. Within experimental error (~20%) both values are the same.

Figure 4a is the SEM image of a photonic crystal fibre with a (68 ± 6) μm deep hole. Re-deposition of the sputtered silica is observed on the internal walls and edge of the hole – it appears on one side only indicating that the FIB is not perpendicular to the fibre surface. The formation of a periodic-like structure ($\Lambda$ ~ 0.7 μm) is also noticed. This is most likely due to interference effects between the ion beam and the plasma discharged from the hole. Further, the ablated material and subsequent reductions in penetration through this at greater depths by the ion beam may be responsible for a comparable milling rate between the structured fibre and a solid fibre. This ablated material may be removed by adopting an additional etching procedure as has been applied with femtosecond lasers[8].

*Fibre (2):* The 6-hole fibre has a thin silica layer between the outside surface and the air holes (~7 μm) leading to faster milling times. No HF etching was necessary. A hole with $\phi$ <15 μm wide was readily milled in 12 mins, allowing direct access into one of the internal channels of the fibre (figure 4b). Given the smaller penetration depth required, there is no visible deposition of ablated material around the hole walls as observed in *Fibre (1)*. The smooth and apparent bulge at the rim of the whole suggests some local melting predominantly along the fibre axis.

In summary, we have demonstrated micromachining of structured optical fibres using a focused ion beam. Results can be applied to all fibres, including different materials where lower energies and more rapid processing can be undertaken, such as in polymer and chalcogenides, fluorides, tellurides and other soft glasses. Hole diameters as small as 10 µm have been demonstrated. In principle, this is limited by the ion beam focal spot which depends on the beam current. Consequently, more complex high-resolution designs will lead to longer milling times.

Cicero Martelli thanks CAPES-Brazil for supporting his scholarship and Giuseppe Tettamanzi for useful discussions. An Australian Research Council (ARC) Discovery Project funded this work.

**Figure Captions:**

Figure 1. a) *Fibre (1):* Photonic crystal fibre , and b) *Fibre (2):* 6-hole step index fibre.

Figure 2. Optical microscope side-view images of milled holes within the photonic crystal fibre samples (depths: $(10 \pm 2)$, $(30 \pm 6)$, $(56 \pm 6)$ and $(70 \pm 6)$ µm from left to right).

Figure 3. Solid-triangles: measured hole depths for FIB conditions ($Ga^+$ ions, 20nA). The solid (experimental best fit) and dashed lines (literature[19]) correspond to the amount of sputtered molecules as function of number of ions for PCF fibre and solid silica respectively.

Figure 4. SEM images of milled holes in a) *Fibre (1):* four ring photonic crystal fibre ($\phi_{diameter}$ ~12 µm, depth ~68 µm), and b) *Fibre (2):* in 6-hole step index fibre ($\phi_{diameter}$ ~15 µm, depth ~7 µm). The channel cavity can be seen below.

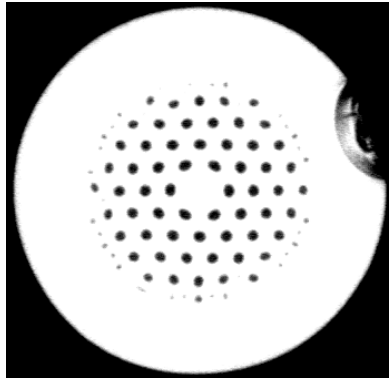 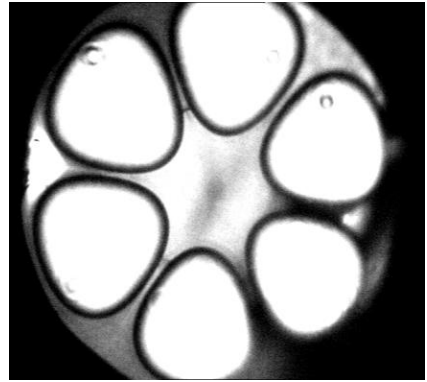

a)            b)

Figure 1. a) *Fibre (1):* Photonic crystal fibre , and b) *Fibre (2):* 6-hole step index fibre.

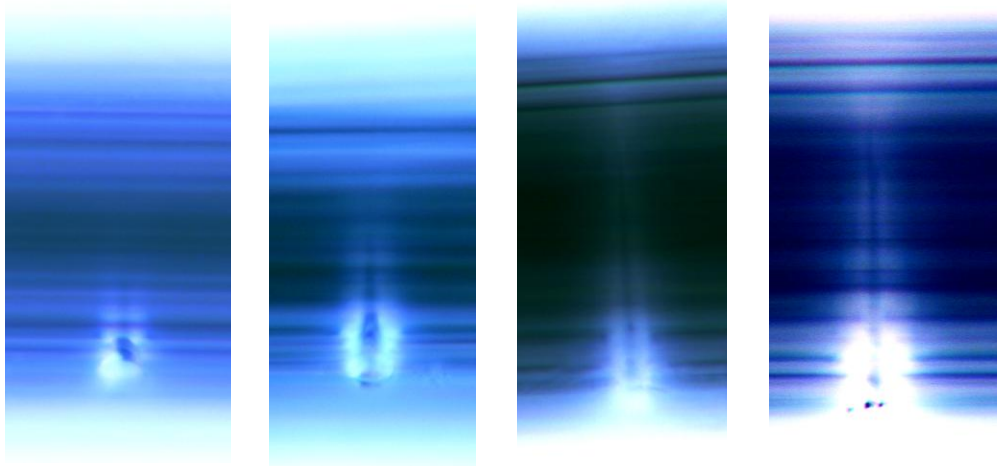

Figure 2. Optical microscope side-view images of milled holes within the photonic crystal fibre samples (depths: (10 ± 2), (30 ± 6), (56 ± 6) and (70 ± 6) μm from left to right).

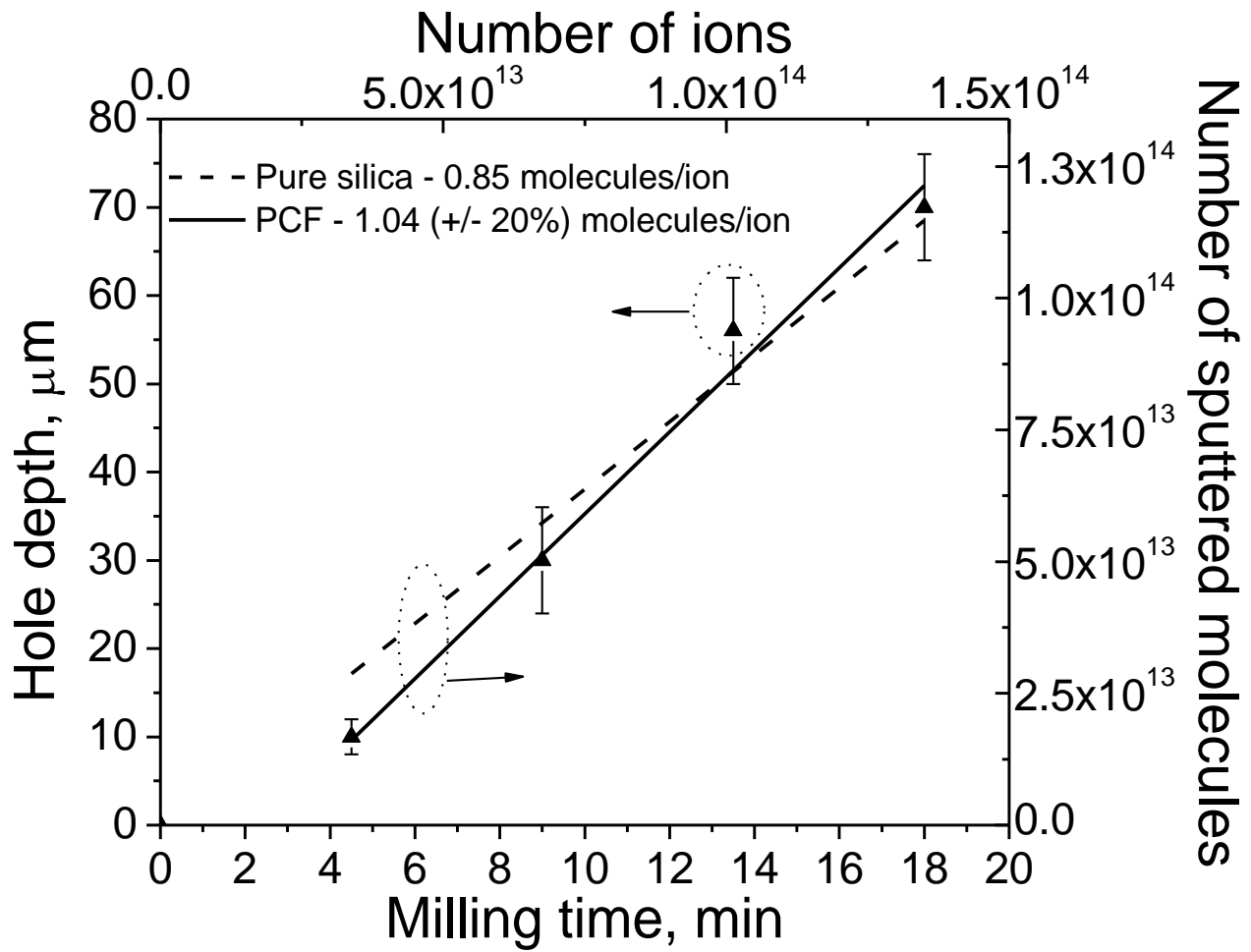

Figure 3. Solid-triangles: measured hole depths for FIB conditions (Ga$^+$ ions, 20nA). The solid (experimental best fit) and dashed lines (literature[19]) correspond to the amount of sputtered molecules as function of number of ions for PCF fibre and solid silica respectively.

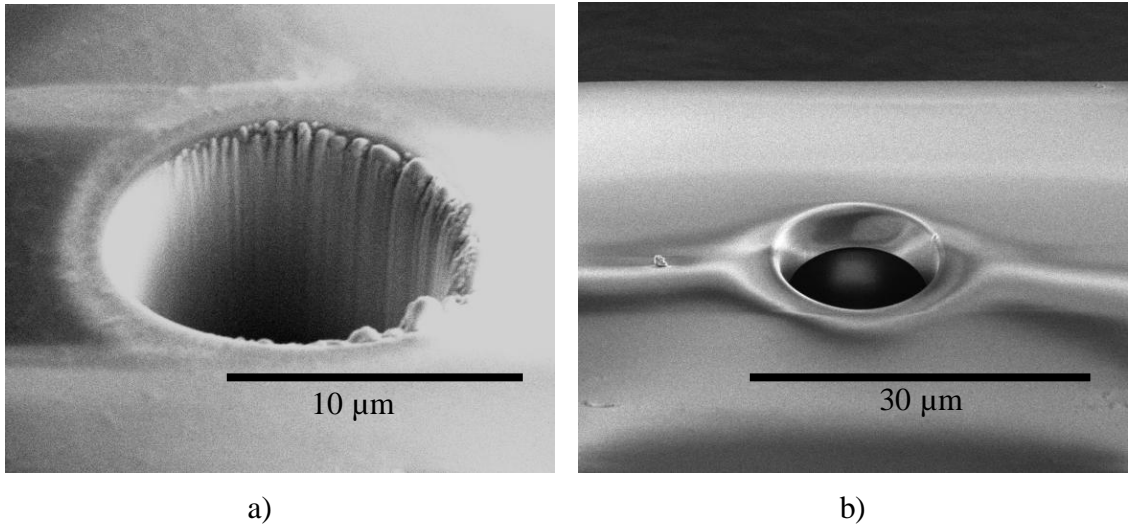

a)                                                     b)

Figure 4. SEM images of milled holes in a) *Fibre (1):* four ring photonic crystal fibre ($\phi_{diameter}$ ~12 μm, depth ~68 μm), and b) *Fibre (2):* in 6-hole step index fibre ($\phi_{diameter}$ ~15 μm, depth ~7 μm). The channel cavity can be seen below.

15